%% file: template.tex
\documentclass{Interspeech}
\usepackage{xcolor}
\usepackage{hyperref}
\hypersetup{
    colorlinks=true,
    linkcolor=blue,
    filecolor=magenta,      
    urlcolor=cyan,
}

\interspeechcameraready

\title{Towards Human-like Multimodal Conversational Agent \\ 
by Generating Engaging Speech}

\author[affiliation={}]{Taesoo}{Kim}
\author[affiliation={}]{Yongsik}{Jo}
\author[affiliation={}]{Hyunmin}{Song}
\author[affiliation={}]{Taehwan}{Kim}

\affiliation[nocounter]{Artificial Intelligence Graduate School}{UNIST}{Republic of Korea}
\email{\{taesoo0630, josik, hyunminsong, taehwankim\}@unist.ac.kr}

\usepackage{comment}

\begin{document}

\maketitle

% the abstract here must exactly match the abstract entered into the paper submission system
\begin{abstract}    
    Human conversation involves language, speech, and visual cues, with each medium providing complementary information. For instance, speech conveys a vibe or tone not fully captured by text alone. While multimodal LLMs focus on generating text responses from diverse inputs, less attention has been paid to generating natural and engaging speech. We propose a human-like agent that generates speech responses based on conversation mood and responsive style information. To achieve this, we build a novel MultiSensory Conversation dataset focused on speech to enable agents to generate natural speech. We then propose a multimodal LLM-based model for generating text responses and voice descriptions, which are used to generate speech covering paralinguistic information. Experimental results demonstrate the effectiveness of utilizing both visual and audio modalities in conversation to generate engaging speech. The source code is available in \url{https://github.com/kimtaesu24/MSenC}
\end{abstract}

\input{intro}
\input{data}

\input{model}
\input{exp}
\input{conc}

\clearpage

\section{Acknowledgements}
This work was supported by Institute of Information \& communications Technology Planning \& Evaluation (IITP) grant funded by the Korea government (MSIT) (No.RS-2022-II220608/2022-0-00608, Artificial intelligence research about multimodal interactions for empathetic conversations with humans, No.IITP-2025-RS-2024-00360227, Leading Generative AI Human Resources Development \& No.RS-2020-II201336, Artificial Intelligence graduate school support(UNIST)) and the National Research Foundation of Korea(NRF) grant funded by the Korea government(MSIT) (No. RS-2023-00219959).

\bibliographystyle{IEEEtran}
\bibliography{mybib}

\clearpage
\input{appendix}

\end{document}

%% file: intro.tex
\section{Introduction}

% 1. What is the conversation? Why this task is important?
In real life, people communicate through multimodal signals and interpret others' non-verbal cues. 
This highlights the importance of multimodal understanding, where words, facial expressions, and speech tones contribute to interpreting meaning. 
Furthermore, individuals adapt their responses based on these cues, not only in what they say but also in how they express it, reflecting subtle differences in tone, emphasis, and delivery.

% 2-1. How can we make multimodal conversation agent? (Question Answering systems + TTS)
% 2-2. What is the problem with these approaches? (only generate text output / cannot reflect paralingual information)
Recently, communication with machines has made significant progress due to the remarkable success of large language models (LLMs), which demonstrate a high level of common knowledge \cite{achiam2023gpt}. 
For instance, text-based QA systems \cite{dubey2024llama,  alpaca}, 
visual QA systems \cite{liu2024visual, liu2023improved}, 
video QA systems \cite{wang2024qwen2, li2024llava}, 
audio-video QA systems \cite{zhang-etal-2023-video} 
can interpret text, video, and audio inputs. Despite these advances, these models are currently only capable of generating text responses. 
There have also been attempts to generate other modalities using LLMs. These models try to retain the semantic information of the input 
but often struggle with cross-modal consistency \cite{wu2023next, tang2024codi}
or lose acoustic details in generated speech due to the usage of speech tokens \cite{zhan2024anygpt, zhang2023speechgpt}.
Integrating a text-to-speech (TTS) module with LLMs is a straightforward approach that enables effective speech interaction. However, current TTS modules \cite{shen2023naturalspeech, li2024styletts, lee2023hierspeech++} are inadequate for human-like communication that considers paralingual information reflecting the mood of communication.

\begin{figure}[t]
\centering
\includegraphics[width=1\columnwidth]{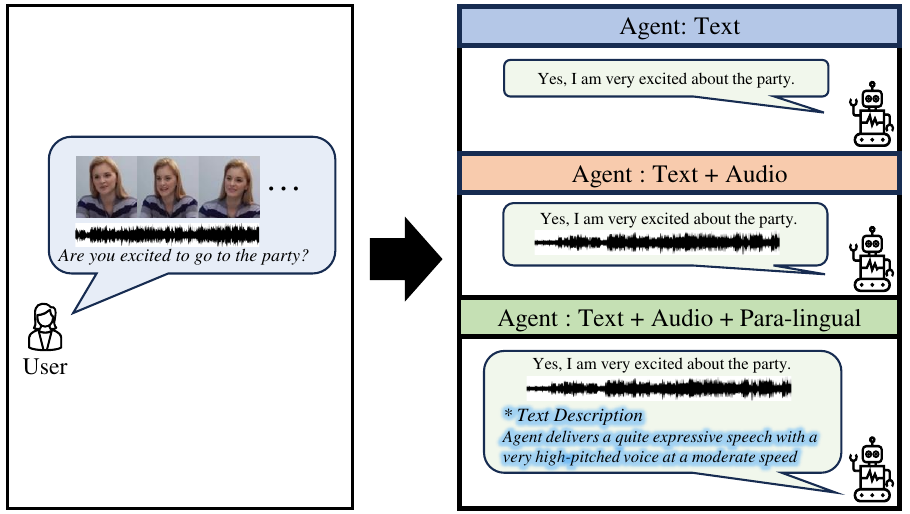}
\vspace{-2em}
\caption{A conversational agent with (Top) text, (Middle) text and audio, (Bottom) text, audio, and paralinguistic signals.}
\label{fig1}
\vspace{-1.5em}
\end{figure}

% 3. How can we handle them? (Introduce new dataset + the core idea)
Developing the proposed conversational agent requires a large-scale corpus of multimodal interactive conversation data. 
However, this presents a significant challenge due to the limitations of existing datasets, which are often constrained by their smaller size or the lack of certain modalities, such as audio. To overcome these limitations, we present a new dataset, \emph{MultiSensory Conversation (MSenC)} dataset. Our dataset is a carefully curated collection of about 31,000 utterances extracted from daily conversation YouTube videos. 
The creation of such a conversational model relies on exposure to this diverse range of multimodal conversation dataset and requires the seamless integration of textual, visual, and acoustic elements. To comprehend multimodal information in conversations, we adopt the BLIP-2 \cite{li2023blip} approach to ensure efficient cross-modal training. 
Finally, to communicate with paralinguistic components derived from the overall communication mood, we utilize LLM and instruction tuning which can guide our model in generating voice descriptions.
By generating responsive voice descriptions that consider the conversation history, we can enhance the naturalness and contextual appropriateness of dialogue systems, as illustrated in Figure~\ref{fig1}.

% 4. Contribution
The contributions of our work can be summarized as follows:
To the best of our knowledge, we are the first to study a dialogue model incorporating para-lingual output in responses. We generate a response with paralinguistic information reflecting multimodal factors in conversation.
We introduce the MultiSensory Conversation dataset, which will be publicly available to advance research in multimodal conversational agents.
Our model effectively utilizes both visual and auditory modalities, producing contextually appropriate speech responses, as validated by both quantitative metrics and qualitative assessments.

%% file: data.tex
\section{MSenC Dataset}
Most existing multimodal conversation datasets \cite{ephrat2018looking, wang2020mead, chu2018face} focus on single-speaker utterances and lack comprehensive multimodal features. Another dataset \cite{park-etal-2024-lets} provides facial images and audio in communication but shows fixed spatial information such as a green screen background. Notably, a dataset \cite{poria2018meld} covers many requirements but is designed for emotional analysis, leading to imprecise audio splitting and background noise from the audience. 
To effectively communicate in a more human-like way, a dataset that encompasses conversing with human faces, rich visual context, and high-quality voice is desirable. 

\begin{figure}[h]
\includegraphics[width=1\columnwidth]{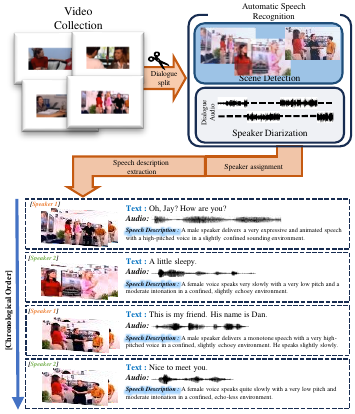} 
\vspace{-1.5em}
\caption{The illustration depicts the creation process of the MultiSensory Conversation dataset.}
\label{fig2}
\vspace{-1.5em}
\end{figure}

To address these limitations, we have taken the initiative to develop our novel dataset, the MultiSensory Conversation (MSenC) dataset depicted in Figure~\ref{fig2}. This dataset, sourced from YouTube and designed for daily conversation, ensures that there is no background music and the spoken English is clear and high quality, 
preventing overlaps, disfluencies, and non-speech vocalizations.
Each scene includes a person, presenting natural conversations with rich visual and auditory elements that help enhance the contextual understanding of dialogue situations.
These videos offer a diverse range of voice features and interactions across various scenarios and contexts, crucial for developing robust models. The total video length is 21.5 hours. The average duration of an utterance is 2.46 seconds.

\subsection{Preprocessing}
\subsubsection{Dialogue Split}
Manually segmenting over 36 hours of videos based on speech is a challenging task. However, it is crucial to carefully check for any unnecessary parts to ensure the content is suitable for learning conversations. So we proceeded to manually segment and filter out the dialogue by human to ensure fairness and accuracy following criteria: 1) When multiple dialogues occurred within the same context such as the individuals involved changed. 2) When the scene transitioned to a different setting during the conversation.

\subsubsection{Utterance Split}
To efficiently segment the dialogue into individual utterances, we can employ speaker diarization, which identifies speakers in an audio recording and assigns timestamps to their speech. While this approach faces challenges, such as difficulty in accurately distinguishing speakers and a tendency to overly fragment.
To address these issues, we incorporated automatic speech recognition (ASR) with timestamp capabilities. In our approach, we utilized a pre-trained ASR model\footnote{https://huggingface.co/distil-whisper/distil-large-v3} that trains OpenAI's Whisper-large-v3 \cite{radford2023robust} on English-only data, providing more accurate and faster inference speeds. However, since this model is trained for audio clips up to 25 seconds long, it struggles to accurately timestamp longer clips. To overcome this, we applied a scene detector\footnote{https://github.com/Breakthrough/PySceneDetect} to divide longer audio into shorter clips. For clips that are still longer than 25 seconds, we employed speaker diarization\footnote{https://huggingface.co/pyannote/speaker-diarization-3.1}. This method allowed us to more effectively segment the entire video into distinct speech units, each corresponding to individual speakers.

\subsection{Metadata Processing}
\subsubsection{Speaker Assignment}
We assign a speaker ID to each video clip according to dialogue units. While speaker diarization is the desirable method for indexing speakers to utterances, it has limitations in performance. We take an alternative approach to address this limitation: cluster the speech embeddings. Figure~\ref{app_fig4}, located in the appendix, illustrates the overall structure of our approach. We obtain speech embeddings from each speech clip using WeSpeaker \cite{wang2023wespeaker}, a model focused on learning speaker embedding, particularly for speaker verification tasks. 
Specifically, we use HDBSCAN \cite{mcinnes2017hdbscan}, an algorithm capable of handling varying densities and does not require predefining the number of clusters. This flexibility is particularly important in our environment, where the number of participants is variable.
We employ cosine distance as the distance measure since most speaker verification systems utilize cosine similarity for evaluation. 
To assess the quality of our process, we calculated the accuracy of our method by manually labeling 20 dialogues, which included a total of 602 samples, resulting in an accuracy of 95.49\%. This demonstrates the effectiveness of our speaker assignment method.

\subsubsection{Speech Description}
Since our goal is generating engaging speech, we extracted speech descriptions that accurately capture the characteristics of the speech. Parler-TTS \cite{lyth2024natural} is a text-to-speech system that transforms text into speech, incorporating detailed paralingual descriptions.
This system provides methods~\footnote{https://github.com/huggingface/dataspeech} for extracting annotations of speaking style and generating audio descriptions derived from these annotations.

For processing the MSenC dataset, we extract annotations including gender, pitch, speech monotony, speaking pace, and reverberation.
Especially for gender, which is needed for generating speech descriptions but cannot be derived directly, we perform gender recognition~\footnote{https://huggingface.co/alefiury/wav2vec2-large-xlsr-53-gender-recognition-librispeech} from raw speech, achieving an F1 score of 99.93\%.
Subsequently, the LLaMa-3 \cite{dubey2024llama} generates natural language descriptions that effectively convey the conversation mood based on these annotations.

%% file: model.tex
\begin{table*}[t]
\centering
\begin{tabular}{l|cccc|cccc}
\hline
% \multicolumn{1}{c|}{}          & \multicolumn{8}{c}{Datasets}                                                                                                                                  \\ \hline  
\multicolumn{1}{c|}{}          & \multicolumn{4}{c|}{MSenC}                                                    & \multicolumn{4}{c}{MELD  \cite{poria2018meld}}                                                      \\ \hline
\multicolumn{1}{c|}{Modality}  & B@1               & B@3               & METEOR            & ROUGE             & B@1               & B@3               & METEOR            & ROUGE             \\ \hline
Text                           & 12.30             & 4.11              & 5.81              & 11.90             & 7.99              & 1.60              & \underline{4.47}  & 8.09              \\
Text + Audio                   & 12.96             & \underline{4.82}  & 6.27              & 11.83             & \underline{9.10}  & \underline{2.11}  & 4.35              & \underline{8.24}  \\
Text + Video                   & \underline{14.62} & 4.78              & \underline{6.63}  & \underline{13.38} & 5.62              & 1.00              & 2.53              & 4.03              \\
Text + Audio + Video           & \textbf{15.11}    & \textbf{5.25}     & \textbf{6.89}     & \textbf{14.12}    & \textbf{10.23}    & \textbf{2.19}     & \textbf{4.74}     & \textbf{9.88}     \\ \hline
\end{tabular}
% \vspace{-0.4em}
\caption{Ablation study on different modalities across two datasets. The text-only modality model represents a pure LLM that has been fine-tuned with each dataset. "B" stands for BLEU score.}
\label{tab1}
\vspace{-2.0em}
\end{table*}

\begin{figure}[t]
\centering
\includegraphics[width=1\columnwidth]{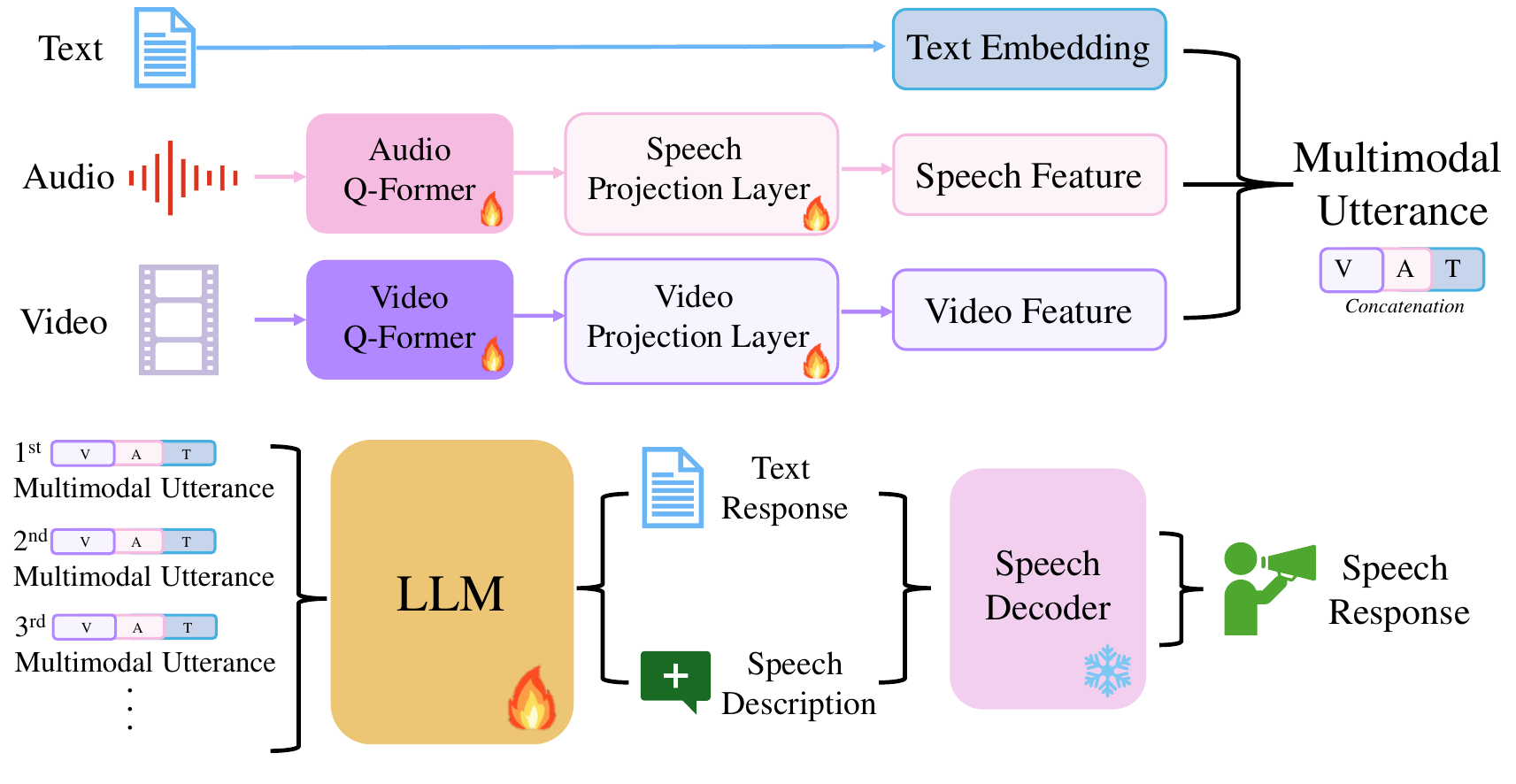}
\vspace{-2.5em}
\caption{Overview of model architecture. The multimodal utterances are composed of text, audio, and video features. Then LLM generates text response and speech description with them.}
\label{fig3}
\vspace{-1.5em}
\end{figure}

\section{Method}

We develop a model capable of processing multiple modalities and generating engaging speech within a large language model. Figure~\ref{fig3} shows an overview of our architecture. Our model takes a set of images, audio, and text as a single utterance input and generates a responsive textual sentence output along with a voice description. We denote our dataset as $D = \left\{{d^a, d^v, d^l}\right\}$ where $a$ represents the acoustic modality, $v$ the visual modality, and $l$ the linguistic modality, with $d$ indicating dialogue. Each dialogue consists of a set of utterances. Let $d^m=\left\{ {u_1^m, u_2^m, ...,u_t^m, u_{t+1}^m} \right\}$ represent
a single dialogue, where $t$ denotes time step, and $m$ presents a certain modality.

\subsection{Multimodal Understanding}
We utilize Q-Former following Video-LLaMA \cite{zhang-etal-2023-video}, where video and audio modalities are processed using Q-Formers with the same structure as Blip-2 \cite{li2023blip}. This Q-Former has demonstrated strong performance by enhancing computational efficiency and model stability. It produces fixed-size features regardless of the length of input video or audio, which simplifies the integration of multimodal data and ensures consistent input sizes for subsequent processing. To initialize the Q-Former, we adopt the pretrained model from Blip-2 \cite{li2023blip}. 
These models are then fine-tuned to enable our model to capture visual context and auditory information effectively.

For single utterance $u^m_{t} = \left\{u^a_{t}, u^v_{t}, u^l_{t}\right\}$, the video and audio inputs are processed separately.
For video sampling, we uniformly extract three frames per second and consider the list of images as a conversation scene. In contrast, the audio processing takes the entire speech as input. While video sampling is conducted to reduce redundant information and improve efficiency, the same method cannot be applied to audio due to significant information loss. 
The sampled data are applied to their respective Q-Former and then projected into the text embedding space of a large language model through the linear projection layer. The resulting features are concatenated with those from other modalities and used as utterance representations.

\subsection{Speech Description Generation}
The processed utterances are combined with the conversation history $\left\{u^m_1, u^m_2,..., u^m_{t-1}, u^m_{t}\right\}$ and fed into the LLM to understand the context comprehensively.
To provide richer communication, we train our model to incorporate not only linguistic information but also paralinguistic cues by describing voice. This is achieved through instruction tuning, a new process where voice descriptions are created after the language model generates responses. 
Ultimately, the model generates the response and speech description in textual format.
We also provide instructions to specify which speaker delivers each utterance, enabling the model to respond or continue the previous utterance.

\subsection{Training Loss}
We use a target response sentence paired with its corresponding audio description. The cross-entropy loss is then computed between the target $\left\{{u}^l_{t+1}, {desc}_{t+1}\right\}$ and the model output $\left\{\hat{u}^l_{t+1}, \hat{desc}_{t+1}\right\}$, as illustrated in Equation~\ref{eq:cross_entropy}, using the concatenation operation denoted by $\mathbin\Vert$.
\begin{equation}
  \label{eq:cross_entropy}
  Loss = CE({u}^l_{t+1} \mathbin\Vert desc_{t+1}, \hat{u}^l_{t+1} \mathbin\Vert \hat{desc}_{t+1})
\end{equation}
Furthermore, we fine-tune the LLM backbone with parameter-efficient fine-tuning \cite{hu2021lora} to specialize the model specifically for generating paralinguistic descriptions in the conversation.

%% file: exp.tex
\section{Experiment}
\subsection{Experimental Setup}
In our experiments, we evaluate our model on the MSenC dataset and MELD \cite{poria2018meld}. We extract visual features using CLIP-VIT \cite{radford2021learning}.
The acoustic features are obtained from WavLM \cite{chen2022wavlm}. 
We utilize Mistral-7B \cite{jiang2023mistral} as our LLM backbone and utilize Parler-TTS \cite{lyth2024natural} as our speech decoder. 
The experiments were conducted using a single NVIDIA A100 80G GPU. We use a batch size of 6 and training has spent 30 hours.

\subsection{Text Analysis}
\subsubsection{Modality Ablation}
Since our model processes multimodal input, we assess how each modality impacts performance by examining changes in metrics.
METEOR \cite{banerjee2005meteor} measures not just word overlap but also semantic meaning, and ROUGE \cite{lin2004rouge} measures coherence and flow. 
We calculate the score only on the text response, excluding the voice description.
Table~\ref{tab1} shows the impact of audio and video modality processed through the Q-Former. 
According to the MSenC dataset result, incorporating additional modalities enhances the quality of the responses, indicating a positive effect on multimodal understanding. Specifically, combining audio, video, and text yields the highest performance, suggesting that responses are more appropriate, contextually accurate, and natural-sounding.
Similar results are observed with the MELD \cite{poria2018meld}, where incorporating audio and video inputs also results in the highest performance.

\begin{figure}[t]
\centering
\includegraphics[width=1\columnwidth]{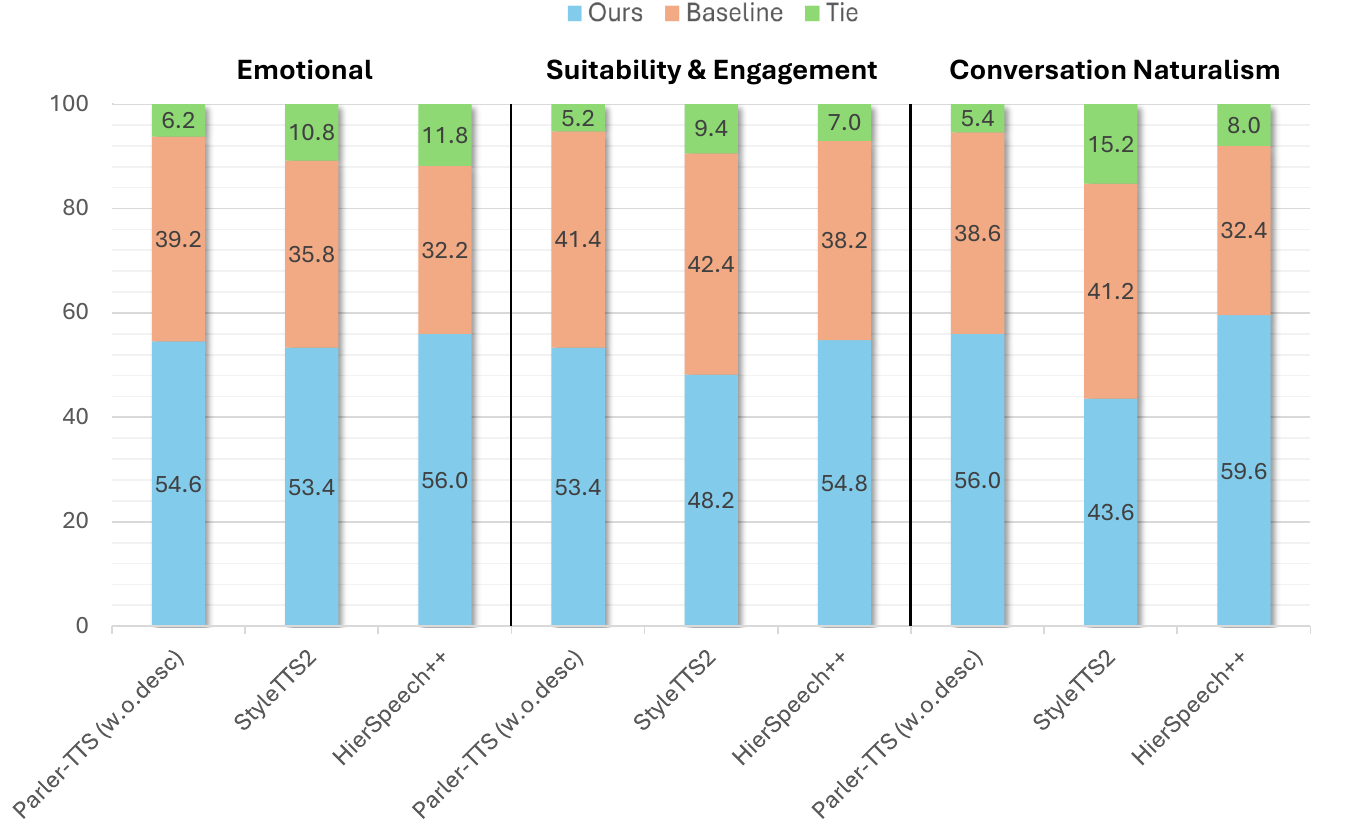}
\vspace{-2em}
\caption{User study results on the MSenC test dataset.}
\label{fig4}
\end{figure}

\begin{table}[t]
\centering
\begin{tabular}{c|c}
\hline
\multicolumn{1}{c|}{Model} & Accuracy \\ \hline
Ours                       & \textbf{15.10}\%   \\
Parler-TTS \cite{lyth2024natural} (w.o. description)    & 11.20\%   \\
StyleTTS2 \cite{li2024styletts}                  & 13.72\%   \\
HierSpeech++ \cite{lee2023hierspeech++}               & 12.54\%   \\ \hline
\end{tabular}
\vspace{0.4em}
\caption{Result of evaluating emotional continuity in conversations on the MSenC test dataset.}
\label{tab2}
\vspace{-2.4em}
\end{table}

\subsection{Speech Analysis}
\subsubsection{User Study} 
We conducted a human evaluation how to assess the conversational speech style generated by our system contributes to expressiveness, focusing on the impact of paralinguistic information.
We used Amazon Mechanical Turk for the assessment, which involved 5 judges and 100 generated samples. The conversational history was limited to a maximum of five entries, presented through video content. The evaluation focused on three criteria: emotional, suitability \& engagement, and conversational naturalness.
We compared our system with StyleTTS2 \cite{li2024styletts}, which applies a suitable speaking style to the input text; HierSpeech++ \cite{lee2023hierspeech++}, a zero-shot speech synthesis framework that enhances robustness and expressiveness; and Parler-TTS \cite{lyth2024natural}, which can generate speech from a natural language prompt, and in our comparison, we evaluated it without such a prompt. Comparisons are made against their official checkpoints.
The speech sample was generated from the test set of the MSenC dataset. As in Figure~\ref{fig4}, our model shows superior results on every criterion, which demonstrates the effectiveness of our approach to generate expressive speech reflecting conversation mood.

\subsubsection{Emotion Classification}
We classify the emotions of each utterance and calculate accuracy based on whether the emotions match. The process assumes that if an utterance's emotion aligns with the previous one, it is considered empathetic, highlighting continuity in dialogue.
Using the MSenC dataset and a pretrained speech emotion classification model\footnote{https://huggingface.co/ehcalabres/wav2vec2-lg-xlsr-en-speech-emotion-recognition}, we classify one of eight emotions: angry, calm, disgust, fearful, happy, neutral, sad, or surprised. Our model generates each speech and then compares it with a prior speech to assess emotional consistency. Table~\ref{tab2} demonstrates that our model outperforms the baseline models in maintaining consistent emotional expression across the conversation.

\subsection{Qualitative Analysis}
\begin{figure}[t]
\centering
\includegraphics[width=1\columnwidth]{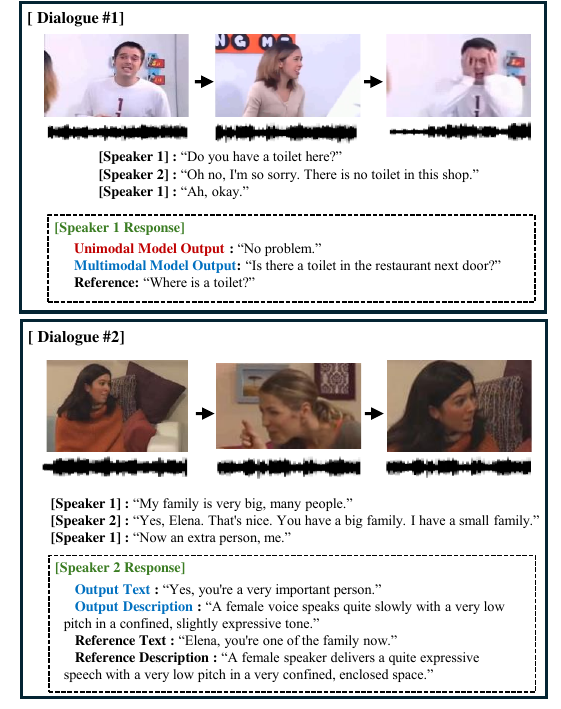} 
\vspace{-2.4em}
\caption{Qualitative analysis samples evaluated on the MSenC test dataset.}
\label{fig5}
\vspace{-2.0em}
\end{figure}

Our previous analysis highlights our model's ability to understand multimodality and determine how the text should be spoken. However, it is worth noting that metrics alone might not capture the full essence in an open-domain scenario. 
Consequently, we present a comparative analysis illustrated in Figure~\ref{fig5}. Dialogue \# 1 compares the multimodal model with the text-based unimodal model. The speaker's gestures in the video and tone of voice in the audio convey an urgent situation. These additional modalities enable our model to generate more contextually appropriate responses. In Dialogue \#2, the output shows that our model generates speech descriptions with similar characteristics to the reference, including pace, pitch, and tone. This leads to more engaging and contextually suitable speech responses. 
Overall, our model has a better understanding of multimodal inputs, generating engaging responses that closely match the context and improve relevance.

%% file: conc.tex
\section{Conclusion}
We study a dialogue model with visual and audio inputs from a speaker to generate engaging speech. 
We propose a novel dataset that is specifically curated for training such models.
Then we introduce a novel conversation model that outperforms the baselines in experiments and thus shows its effectiveness. 
Our model cannot replicate a speaker's exact voice from historical recordings, but this does not affect inference since the agent consistently uses a single voice. 
We believe our approach contributes to more natural and human-like conversation, and our proposed dataset may further promote subsequent research.

%% file: appendix.tex
\clearpage
\appendix

\section{Implementation Details}
\label{sec:implementation}

We utilize Mistral-7B~\cite{jiang2023mistral} as our LLM backbone. We train our model with the following hyperparameters. We use a batch size of 6 and Adam optimizer with learning rate of 5e-5 and learning rate decay of 0.98. The video padding size is 50, audio padding size is 800. This size made the same number of utterances in a single dialogue history. We sample the video data, capturing frames at a rate of three per second for each utterance, while the audio remains unsampled. We set the maximum input length for LLM as 800 which can cover about 10 multimodal histories. They are truncated from the oldest history to prioritize focusing more on the latest utterance.
Finally, we tuned the number of epochs on validation data and chose epoch 10.
Our experimental environment was conducted using a single NVIDIA-A100 80G GPU. Training has spent 30 hours.\\

\section{MSenC Dataset Details}
\label{sec:data_count}
In this section, we show further details of the new MSenC dataset. The statistics are presented in Table~\ref{tab3}. To summarize, we divided the video content into 1,120 dialogues and 31,409 utterances. The total video length is 21.5 hours. The average duration of an utterance is 2.46 seconds. The histograms of video durations and word count can be found in Figure~\ref{duration_hist}. Note that many videos begin with greetings such as "Hello" or "Good Morning", which contribute to a higher word count due to their conciseness. Additionally, we evaluated gender bias within our dataset in Table~\ref{tab4}. This involved analyzing the distribution of male and female speakers across different conversational contexts. The result shows about 1:1.5 of rate ensuring that the trained conversational system remains equitable and reliable across diverse gender groups.

\begin{figure}[t]
\centering
\includegraphics[width=1\columnwidth]{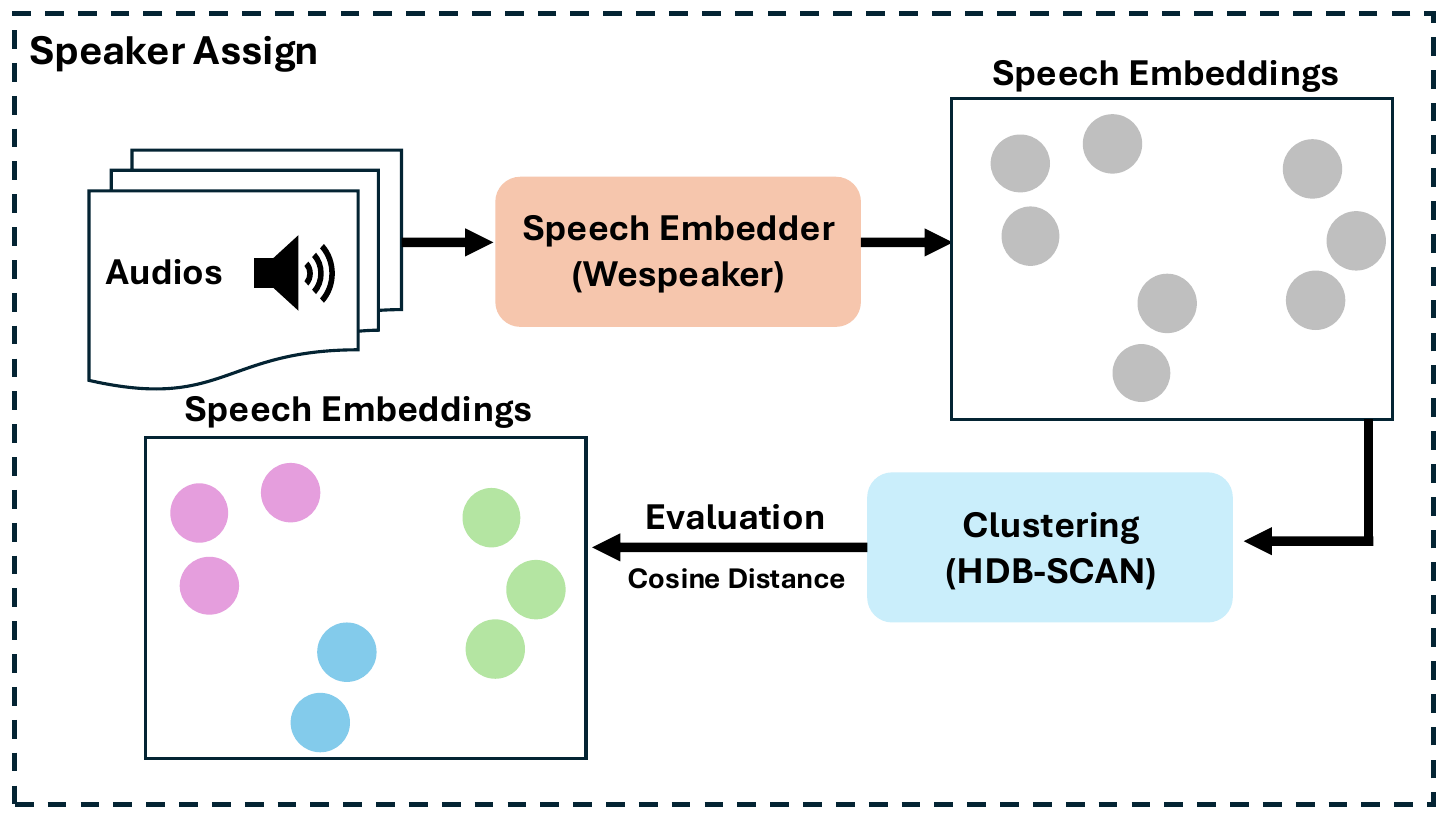} % Reduce the figure size so that it is slightly narrower than the column. Don't use precise values for figure width. This setup will avoid overfull boxes.
\caption{Illustration of speaker assignment pipeline. We obtain speech embeddings and perform clustering.} 
\label{app_fig4}
\vspace{-1em}
\end{figure}

\begin{figure}[h!]
\centering
\includegraphics[width=0.95\columnwidth]{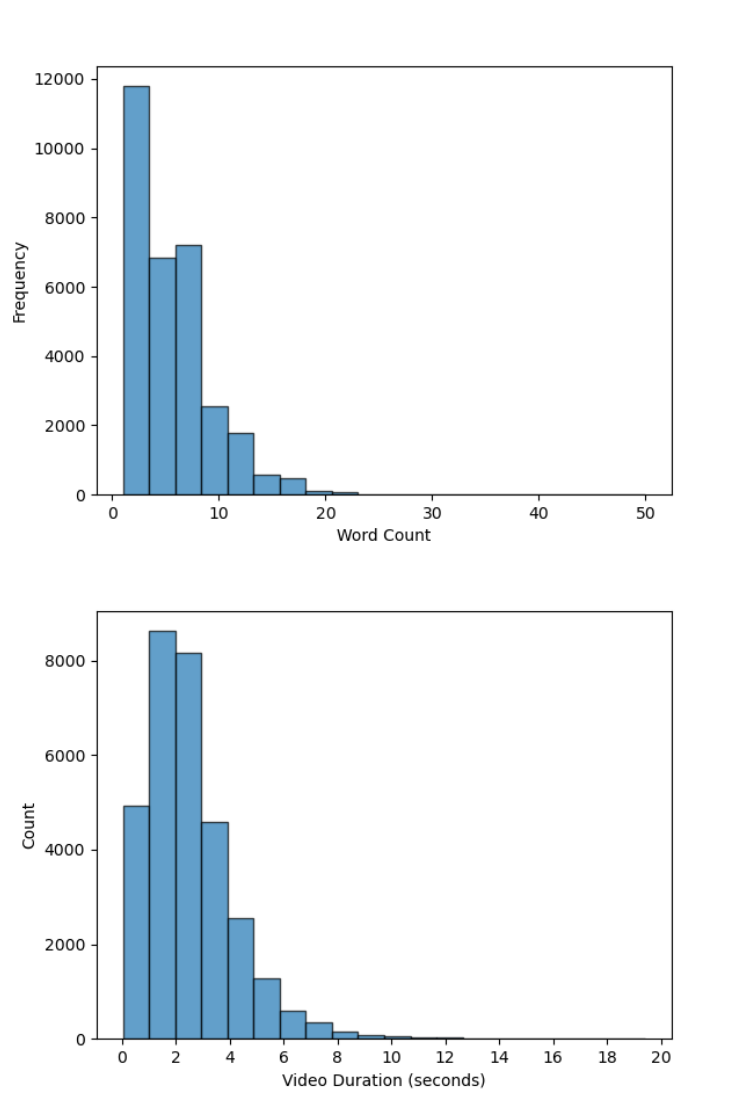} % Reduce the figure size so that it is slightly narrower than the column. Don't use precise values for figure width.This setup will avoid overfull boxes.
\vspace{-1em}
\caption{We report the histogram of video duration in seconds and the histogram of word count in words.} 
\label{duration_hist}
\vspace{-1em}
\end{figure}

For speaker analysis (Figure~\ref{app_fig4}), we extracted speech embeddings from each video clip using a speaker verification model. These embeddings were clustered with HDBSCAN~\cite{mcinnes2017hdbscan}, using cosine distance as the similarity metric.

\begin{table}[h]
\centering
\begin{tabular}{c|ccc|c}
\hline
                     & \textbf{Train} & \textbf{Valid} & \textbf{Test} & \textbf{Total} \\ \hline
\textbf{\# of Dialogue} &  913        & 110           & 97           & 1120            \\     
\textbf{\# of Utterance} & 25624      & 3145          & 2640          & 31409           \\
\hline
\textbf{Duration} & 17.5h      & 2.1h          & 1.8h          & 21.5h           \\ \hline

\end{tabular}
\caption{Statistics of the MSenC dataset.}
\label{tab3}
\vspace{-1em}
\end{table}

\begin{table}[h]
%\resizebox{\columnwidth}{!}{
\centering
\begin{tabular}{c|cc}
\hline
\multicolumn{1}{c|}{} & \textbf{Male} & \textbf{Female} \\ \hline
Train         & 10,267 & 15,357 \\
Validation    & 12,97  & 1,848 \\
Test          & 985    & 1,655 \\ \hline
Total         & 12,549 & 18,860 \\ \hline
\end{tabular}
%}
\vspace{0.4em}
\caption{Table of gender bias within our MSenC dataset.}
\label{tab4}
\vspace{-2em}
\end{table}

\section{Instruction Tuning}
\label{sec:instruction}
\begin{figure*}[t]
\centering
\includegraphics[width=0.9\textwidth]{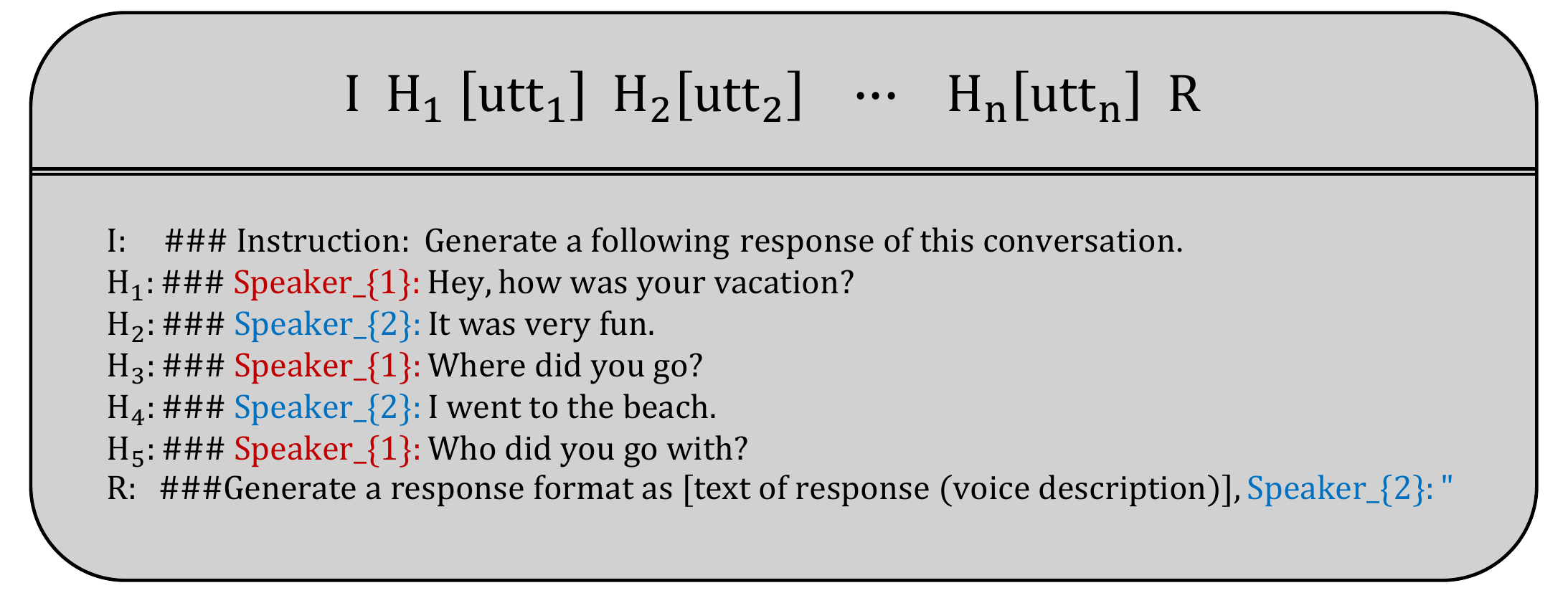}
\vspace{-1em}
\caption{Example of an LLM input with instructions. This sample demonstrates text input for easy understanding, though actual input includes not only text but also integrated text, audio, and video modalities.}
\label{instruction_tuning_sample}
\vspace{1.0em}
\end{figure*}

We’ve introduced instruction tuning in our training process.
We provide comprehensive instruction initially and give speaker ID information for each utterance. At the end, we give additional instruction for generating voice descriptions. Figure~\ref{instruction_tuning_sample} shows a sample of instruction tuning. This sample demonstrates text input for easy understanding, though actual input includes not only text but also integrated text, audio, and video modalities.

\begin{table*}[h]
\centering
\begin{tabular}{l|cccccc|cccccc}
\hline
\multicolumn{1}{c|}{}  & \multicolumn{6}{c|}{MSenC}                                                      & \multicolumn{6}{c}{MELD \cite{poria2018meld}}                                                      \\ \hline
                               & B@1               & B@2               & B@3               & B@4               & METEOR            & ROUGE                 & B@1               & B@2               & B@3               & B@4               & METEOR            & ROUGE          \\ \hline
Ours w.o.ft                    & 13.96             & 7.96              & 5.03              & 3.25              & 6.55              & 12.77      & 5.67              & 2.11              & 0.97              & 0.48              & 2.90              & 4.95                     \\
Ours                           & \textbf{15.11}    & \textbf{8.57}     & \textbf{5.25}     & \textbf{3.35}     & \textbf{6.89}     & \textbf{14.12}       & \textbf{10.23}    & \textbf{4.33}     & \textbf{2.19}     & \textbf{1.21}     & \textbf{4.74}     & \textbf{9.88}  \\ \hline
\end{tabular}
\vspace{0.0em}
\caption{Result of LLM fine-tune on MSenC and MELD dataset.}
\label{tab6}
\vspace{-1.0em}
\end{table*}

\section{LLM Fine-Tuning}
\label{sec:lora}
We evaluated the effect of parameter-efficient fine-tuning on a large language model, with results shown in Table~\ref{tab6}. Fine-tuning led to improved conversational performance compared to the base model. Evaluation was conducted using BLEU-1, BLEU-2, BLEU-3, BLEU-4, METEOR, and ROUGE, where BLEU-2 and BLEU-4 are reported as additional metrics beyond the main results.

\section{Additional Qualitative Samples}
\begin{figure}[t!]
\centering
\includegraphics[width=1\columnwidth]{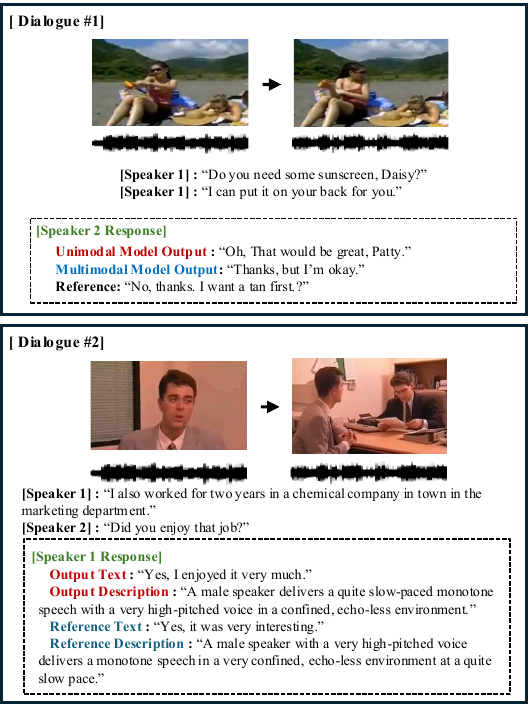} % Reduce the figure size so that it is slightly narrower than the column.
\vspace{-2.0em}
\caption{Qualitative analysis samples evaluated on the MSenC test dataset.}
\label{fig9}
\vspace{-1.0em}
\end{figure}

We provide additional sample of qualitative analysis in Figure~\ref{fig9}.
In Dialogue \#1, we present a comparative analysis of our model's outputs against those of the text-based unimodal model. The output text adapts based on information from the video, resulting in responses that closely match the reference context. In Dialogue \#2, we demonstrate our model's ability to generate speech that conveys how to say the text content. The output shows that our model generates speech descriptions with similar characteristics to the reference, including pace, pitch, and tone. This leads to more engaging and contextually suitable speech responses.

\section{Details of Human Evaluation}
\label{sec:user_study}
\begin{figure*}[]
\centering
\includegraphics[width=\textwidth]
{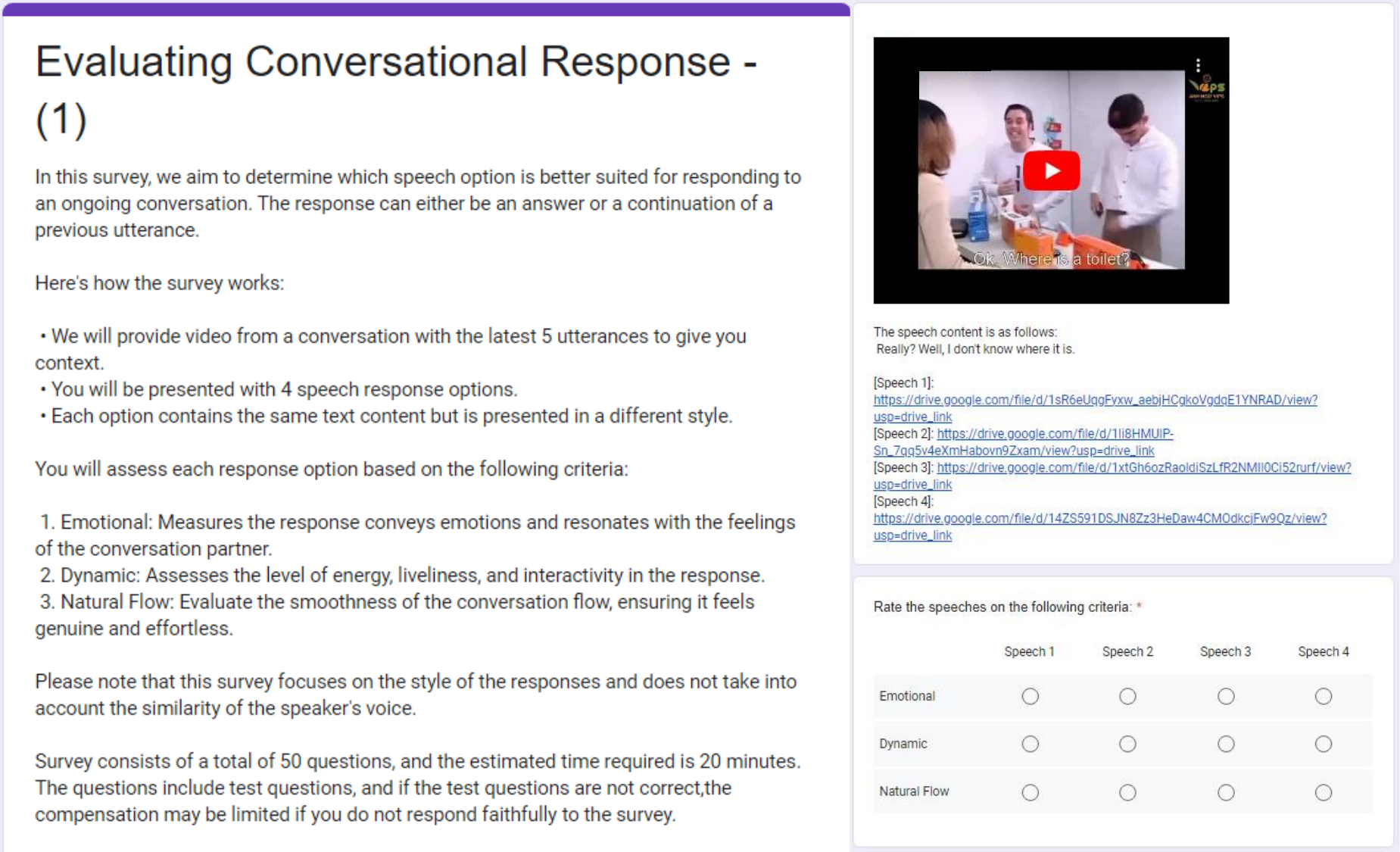}
\caption{Human evaluation template.}
\label{human_evaludation_format}
\end{figure*}

We present our experimental setup as follows: history is limited to a maximum of five entries, and the history is provided through video content. Participants in the experiment are presented with three response options: 'Speech 1', 'Speech 2', 'Tie'. 
% The evaluation criteria for the responses are as follows: "emotional" assesses how well the response conveys emotions and connects with the conversation partner’s feelings, while also measuring the energy, liveliness, and interactivity of the response.
% "suitability and engagement" evaluates how appropriately the response fits within the context and evaluates the level of active and attentive participation in the conversation.
% "conversation naturalism" reflects the overall smoothness of the conversation, ensuring that the interaction feels natural, effortless, and genuine.
The evaluation criteria were: 
\begin{itemize}
    \item \textbf{"Emotional"} assesses how well the response conveys emotions and connects with the conversation partner’s feelings, while also measuring the energy, liveliness, and interactivity of the response.
    \item \textbf{"Suitability \& Engagement"} evaluates how appropriately the response fits within the context and evaluates the level of active and attentive participation in the conversation.
    \item \textbf{"Conversation Naturalism"} reflects the overall smoothness of the conversation, ensuring that the interaction feels natural, effortless, and genuine.
\end{itemize}
We evaluate 100 output samples. This approach guarantees that our evaluation encompasses a diverse range of responses, contributing to the overall reliability of our findings. The template for human evaluation is provided in Figure~\ref{human_evaludation_format}.

\section{Limitations}
Our model cannot replicate a speaker's exact past voice, but this does not hinder inference as the agent consistently uses one voice.
Potential risks include copyright issues with YouTube videos. Because sharing downloaded videos is prohibited, we only release preprocessing code. This ensures users process their own legally obtained data while remaining compliant with copyright regulations.

\clearpage